# Coarse molecular-dynamics analysis of stress-induced structural transitions in crystals

Miguel A. Amat[1], Ioannis G. Kevrekidis[2], and Dimitrios Maroudas[1,*]
[1] *Department of Chemical Engineering, University of Massachusetts, Amherst, MA 01003*
[2] *Department of Chemical Engineering and Program in Applied and Computational Mathematics, Princeton University, Princeton, NJ 08544*

We present a coarse molecular-dynamics (CMD) approach for the study of stress-induced structural transformations in crystals at finite temperatures. The method relies on proper choice of a coarse variable (order parameter, observable), which parameterizes the changes in effective free energy during the transformation. Results are reported for *bcc*-to-*hcp* lattice transitions under pressure. We explore coarse-variable space to reconstruct an effective free-energy landscape quantifying the relative stability of different metastable basins and locate the onset, at a critical pressure, of the *bcc*-to-*hcp* transformation.

Being able to *predict* stable crystalline phases under applied mechanical loading is crucial in understanding polymorphic transitions in crystalline solids. While the molecular-dynamics (MD) method developed by Parrinello and Rahman[1] (PR-MD) has been a major contribution toward capturing solid-solid transformations under stress, it requires (if used on its own) tedious analysis of long transient trajectories near critical points for reasonably accurate predictions of polymorphic transition onsets. To address such challenges, novel developments, namely metadynamics[2-4] and coarse molecular dynamics (CMD)[5] have provided systematic alternatives to mere analysis of conventional MD trajectories.

The purpose of this Letter is to demonstrate the capabilities of CMD as an efficient computational approach to (*i*) locate stable crystalline phases successfully and (*ii*) predict accurately the transformation onset in polymorphic transitions in crystals. Specifically, we focus on determining the loading condition, expressed by a critical pressure $P = P_c$, at which the onset of a transition from a body-centered cubic (*bcc*) to a hexagonal close-packed (*hcp*) phase occurs in a crystal under hydrostatic loading. To validate our approach, we choose as a reference the work of Zhao et al.[6] that has outlined (based on PR-MD simulations and lattice-statics calculations) the stability limits of different Morse-model crystalline phases under hydrostatic loading at low temperature. In particular, we concentrate on the case of the Morse-Ni crystal, where a *bcc* phase is stabilized under compression. For comparison purposes, we mention that, in Ref. 6, the results are reported in terms of stretch factors, $\lambda = (\rho_0/\rho)^{1/3}$, where $\rho_0$ and $\rho$ are the densities of the cubic crystal at zero pressure and at the pressure of interest, respectively.

Our model consists of a cubic supercell with 1458 atoms arranged in a *bcc* lattice. The interatomic interactions are described by a Morse potential properly parameterized for Ni.[6] The equations of motion are obtained through the PR ansatz, where the supercell geometry is described by a matrix, **h**, whose columns correspond to the three vectors, **a**, **b**, and **c**, that define the edges of the supercell. Cell rotations are avoided by setting the subdiagonal elements of **h** to zero; this eliminates three out of the nine degrees of freedom while preserving the full geometrical flexibility of the cell. The equations of motion are integrated using a fifth-order predictor-corrector method with a fine time step (0.4 fs), and the temperature is kept constant using a Berendsen thermostat[7] with a time constant equal to the integration time step. In this study, we report results for a temperature $T = 1$ K; it should be noted that for the pressure-induced transitions of interest, thermal effects are significant even at very low (but finite, $T > 0$ K) temperature close to $T = 0$ K.[6]

Implementation of the CMD approach offers certain major advantages. Specifically, the method provides the means to construct an effective free energy, $G_{eff}$, that can be used to assess the relative stability between phases. These effective free-energy landscapes can be used to determine the transformation onset, expressed by a critical pressure ($P_c$) in the case of pressure-induced structural transitions. The theoretical foundation of CMD has been outlined elsewhere.[5,8-11] Briefly, the method relies on the existence of one or more coarse variables that contain all the necessary information about the long-term dynamics of the system's state; in the case that this study focuses on, we will show that a single such variable, $\psi$, suffices. In addition, the $\psi(t)$ time evolution must be slow and attracting. This means that there exists a one-dimensional attracting manifold, *parameterized* by $\psi$, such that the statistics of the remaining variables of the system become quickly "slaved" to $\psi$ and then evolve slaved to $\psi$. As described elsewhere,[11] the underlying effective Fokker-Planck (FP) equation (for the probability density to find the system at state $\psi$) is identified via short bursts of MD simulation initialized at representative values in the coarse-variable space; each initialization, at $\psi = \psi_0$, is done over many independent copies (twenty-five employed in this study), which are then used to create an ensemble average that is expressed by the brackets, < >. Through the analysis of the averaged time evolution, $<\psi(t)>$, the drift velocity and diffusion coefficient appearing in the FP equation can be obtained as a function of the coarse variable: $\upsilon(\psi) \equiv <[\psi(t+\tau) - \psi(t)]>/\tau$ and $D(\psi) \equiv <[\psi(t+\tau) - \psi(t)]^2>/(2\tau)$ in the limit of $\tau \to 0$, respectively. The information collected from these short MD runs leads to an expression for the equilibrium probability density, which can be integrated to yield $\Delta G_{eff}/kT = -\int \upsilon(\psi')/D(\psi')d\psi'$. The term $\Delta G_{eff}/kT$ encapsulates the logarithmic and integration constant terms obtained in the integration.[11]

[*] Corresponding author. Electronic mail: maroudas@ecs.umass.edu

The change in the supercell geometry as the crystal transforms from, e.g., *bcc* to *hcp*, is reflected in changes among the three angles that span the lattice, $\alpha$, $\beta$, $\gamma$, which are formed by vectors **b** and **c**, **c** and **a**, and **a** and **b**, respectively. Given that changes in these angles are correlated with changes in the state of the crystal, we choose a coarse variable for this transformation as $\psi \equiv (\alpha-90^{o}) + (\beta-90^{o}) + (\gamma-90^{o})$, where a value of $\psi = 0^{o}$ corresponds to a perfect *bcc* lattice. Clearly, the choice of coarse variable is not unique. We have found, e.g., that other scalar variables (combinations of matrix **h** elements) also constitute good coarse-variable choices; we have confirmed that the results of this study do not change with the different coarse-variable choices made.

The implementation of the CMD method is based on two transformations: *lifting* and *restricting*. The former involves initializing the system consistently with a specific value of the coarse variable, $\psi(t = 0) = \psi_0$, while the latter involves monitoring the evolution of the coarse variable. In this particular application, *lifting* is tailored to explore coarse-variable space through deformations covering a range $0^{o} \leq |\psi_0| \leq 5^{o}$. In the lifting scheme that we have implemented, the only angle contributing to the coarse variable, $\psi_0$, is $\alpha$, while $\beta$ and $\gamma$ are kept at 90°. This is achieved by forcing all of the **h** elements to zero except for $h_{11}$, $h_{22}$, $h_{33}$, and $h_{13}$, while imposing an additional constraint that brings the system to the prescribed value of $\psi_0$. The constraints are satisfied at each time step during *lifting* and are implemented by simply modifying the components of the vectors **a**, **b**, and **c** accordingly. This scheme is non-unique and could be implemented through other methods, such as employing Lagrange multipliers. It is emphasized that this procedure still allows for the volume changes required to accommodate the target pressure. The newly lifted states, satisfying $\psi = \psi_0$, are equilibrated at this state for a period of about 9 ps. At the end of this period, all angle constraints are released and the (unconstrained) system is allowed to evolve. During its evolution, $\psi(t)$ is monitored over a time-horizon of 3 ps and used to obtain the values of $\upsilon(\psi_0)$ and $D(\psi_0)$ needed to construct the $G_{eff}$ landscapes at each applied pressure.

Figure 1 shows representative results for the low-temperature evolution of the coarse variable, $\psi(t)$, for various initial conditions $\psi_0$. Figures 1(a) and (b) correspond to reduced pressures of $P$ = 1360 and $P$ = 1340, respectively; $P \equiv P' r_0^3/D$, where $P'$ is dimensional pressure and $r_0$ and D are Morse-potential parameters. At both pressures, initialization at coarse-variable values over the range $2^{o} \leq \psi_0 \leq 5^{o}$ result in trajectories that are invariably "attracted" to the same final state, $\psi(t) = 3.3^{o}$, which corresponds to an *hcp* phase. On the other hand, initializing the system at coarse variables near $\psi_0 = 0^{o}$ ($\psi_0 \leq 0.25^{o}$) yields trajectories that (*i*) fall back to the original *bcc* structure (at higher *P*), or (*ii*) drift away from *bcc* (at lower *P*). Analyses of the entire ensemble of trajectories (25 copies at each choice of $\psi_0$) reveal the existence of three different regions: one attracting region at $2^{o} \leq \psi_0 \leq 5^{o}$ (stable *hcp*), a second region at values of $\psi_0$ near zero (stable/unstable *bcc*) that can be attracting depending on the applied pressure, and a third region at values of $\psi < 2^{o}$ corresponding to intermediate states.

Figure 2(a) shows the $G_{eff}$ landscape corresponding to the conditions of Fig. 1(a). Inset (1) reveals a small thermodynamic-potential well at $\psi$ values near $\psi = 0^{o}$, which corresponds to the stable *bcc* phase. The $G_{eff}$ landscapes were computed at several applied pressure levels within the range $1320 \leq P \leq 1400$.

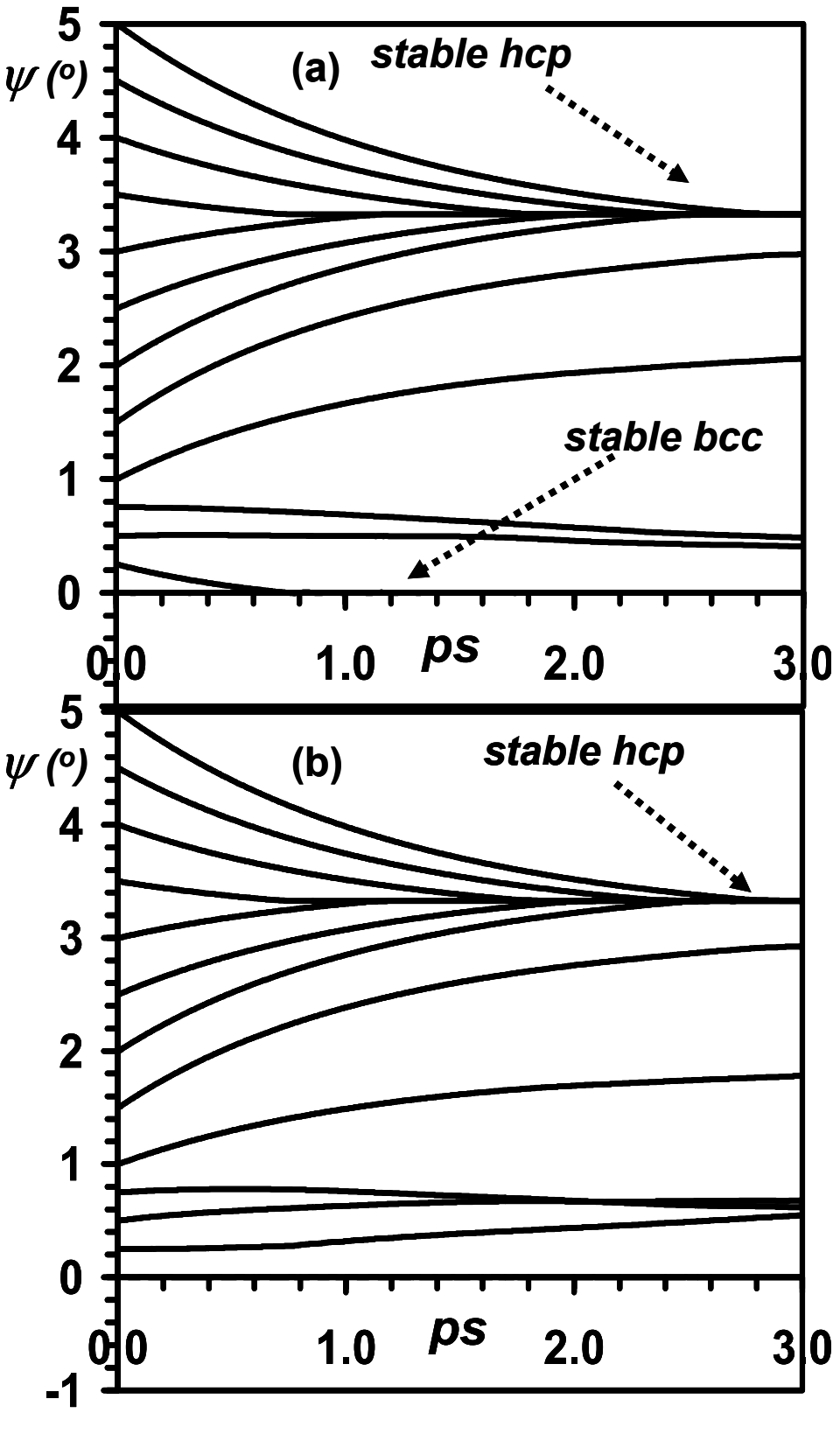

FIG 1. Evolution of the ensemble averaged value of the coarse variable, $\psi(t)$, initialized at different values, $\psi_0$, at reduced pressures of (a) 1360 (above $P_c$) and (b) 1340 (below $P_c$) at $T$ = 1 K.

Based on the relative depth of the wells, it was found that the *hcp* phase is energetically much more favored than the *bcc* one. In addition, it was found that the well corresponding to the *bcc* phase disappears at reduced pressures $P \leq 1345$. The slope $\Delta(\Delta G_{eff}/kT)/\Delta\psi$ also was computed at values near $\psi = 0^o$ and reveals positive/negative values in the presence/absence of the *bcc* well. This slope computation is illustrated in inset (2) to Fig. 2(a) and was used to construct the plot of Fig. 2(b). We relate this change in the sign of the computed slope to changes in the stability of the *bcc* phase: a zero-slope value marks the transformation onset. In Fig. 2(b), the dotted line crossing the ordinate at zero defines two regions: the *bcc* phase is stable above the dotted line and unstable below. Interpolation using the points immediately above and below this dotted line yields a transformation onset of $P_c = 1348 \pm 3$, which corresponds to a stretch factor of $\lambda_c = 0.7492$. This result is in excellent agreement with the values reported in Ref. 6 and validates our CMD approach; the work of Ref. 6 employed conventional MD simulations involving very long MD trajectories (ns-scale runs at each $P$). Finally, we mention that, as shown in the inset to Fig. 2(a), our CMD approach also identifies the transition state, *TS*, of the polymorphic transformation. Analysis of the transition-state configuration and computation of the corresponding kinetic rates for the transformation is beyond the scope of this Letter and will be addressed in detail in a forthcoming publication.

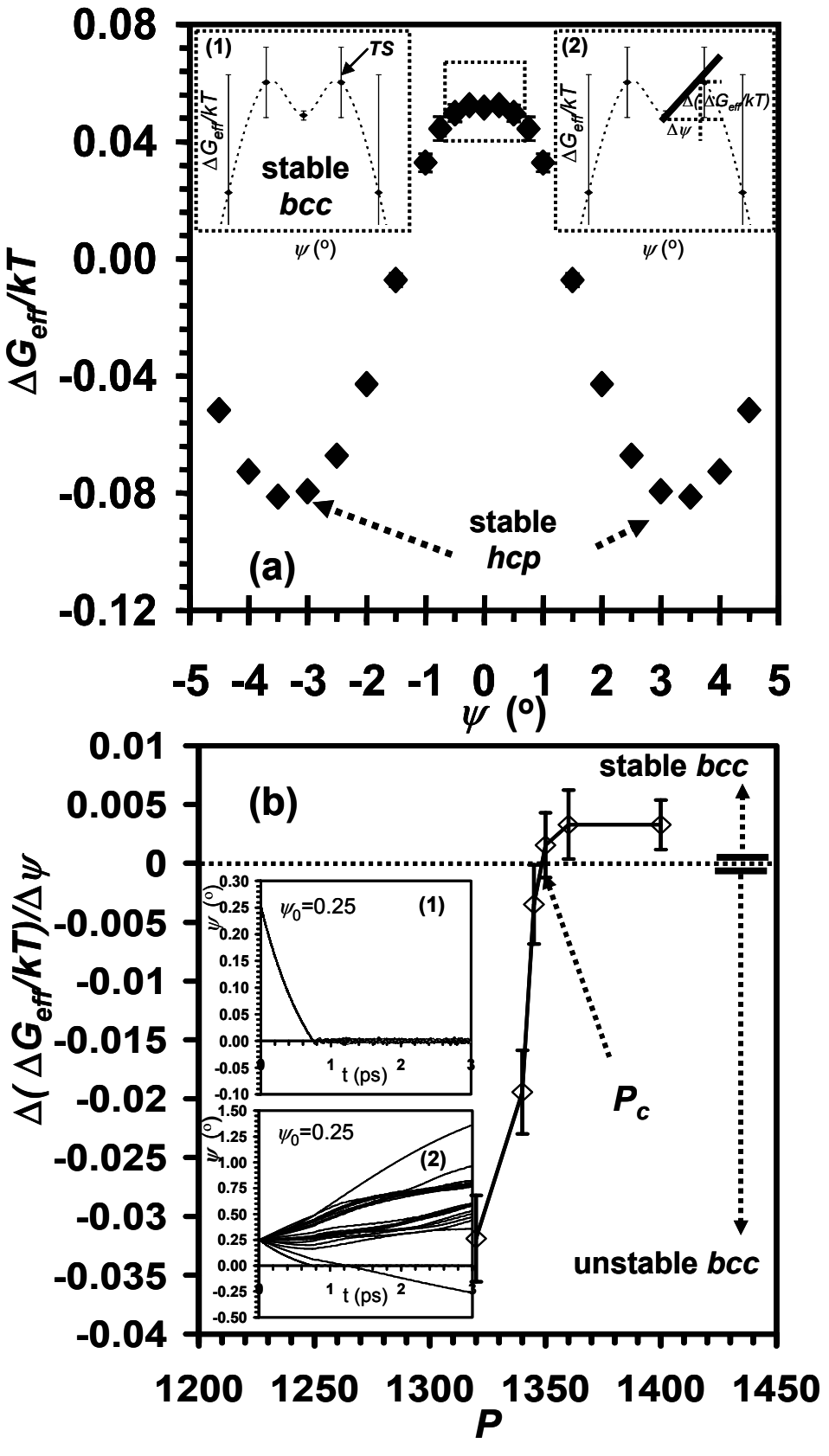

FIG 2. (a) Effective free-energy landscape as a function of the coarse variable $\psi$ at $P = 1360$. Wells centered at $\psi = 0^o$ and $\psi = 3.3^o$ correspond to *bcc* and *hcp* phases, respectively. Insets (1) and (2) correspond to magnifications of the boxed region of the plot with (2) illustrating the slope computation. (b) Slope of $\Delta G_{eff}/kT$ with respect to $\psi$ in the vicinity of $\psi = 0^o$ as a function of $P$. Insets (1) and (2) show representative coarse trajectories of the 25-copy ensemble initialized at $\psi_0 = 0.25^o$ for $P = 1360$ and 1340, respectively. Inset (1) shows a typical attractive evolution toward a stable *bcc* phase (at $P > P_c$), while inset (2) shows a non-attractive evolution at $P < P_c$.

In conclusion, we have demonstrated that implementation of the CMD approach in conjunction with a proper choice of coarse variable succeeds in bracketing and predicting correctly the onset of the *bcc*-to-*hcp* transformation in a model crystal under pressure. This is achieved through construction of effective free-energy landscapes that can be used to assess the relative stability between crystalline phases. In addition, we have shown that implementation of the CMD approach ensures the overcoming of free-energy barriers, which might prevent a material system from crossing between stable states in conventional MD simulations. Although the present study was focused on a particular polymorphic transition under pressure, the approach can be used to study polymorphic transitions under the most general type of applied loading. Finally, we emphasize that the approach is not limited by the choice of force field and can be used with interatomic potentials much more sophisticated than Morse models, as well as with *ab initio* calculated interatomic forces.

This work was supported by the National Science Foundation through Grant Nos. CTS-0205484, CTS-0205584, ECS-0317345, CTS-0417770, and CTS-0613501 and by DARPA. IGK also gratefully acknowledges support from the Guggenheim Foundation.